\documentclass[aps,prl,nobibnotes,twocolumn,superscriptaddress,bibliography]{revtex4-2}

\pdfoutput=1

\usepackage{amsfonts}
\usepackage{mathrsfs}
\usepackage{amsmath}
\usepackage{color}
\usepackage{graphicx}
\graphicspath{ {./pictures/version_8/} }
\usepackage{bm}
\usepackage{amssymb}
\usepackage{xspace}
\usepackage{epstopdf}
\usepackage{dcolumn}
\usepackage{longtable}
\usepackage{multirow}
\usepackage{float}
\usepackage{comment}
\usepackage{chemformula} 

\def\ie{{\it i.e.},\ }

\def\ea{{\it et al.}}

\usepackage[colorlinks=true, letterpaper=true, pdfstartview=FitV, linkcolor=blue, citecolor=blue, urlcolor=blue]{hyperref}

\newcommand{\Rom}[1]{\uppercase\expandafter{\romannumeral#1}}

\makeatother

\begin{document}

\title{Planar Hall Plateau in Magnetic Weyl Semimetals}

\author{Lei Li}
\thanks{These authors contributed equally to this work}
\affiliation{Centre for Quantum Physics, Key Laboratory of Advanced Optoelectronic Quantum Architecture and Measurement (MOE), School of Physics, Beijing Institute of Technology, Beijing, 100081, China}
\affiliation{Beijing Key Lab of Nanophotonics \& Ultrafine Optoelectronic Systems, School of Physics, Beijing Institute of Technology, Beijing, 100081, China}

\author{Chaoxi Cui}
\thanks{These authors contributed equally to this work}
\affiliation{Centre for Quantum Physics, Key Laboratory of Advanced Optoelectronic Quantum Architecture and Measurement (MOE), School of Physics, Beijing Institute of Technology, Beijing, 100081, China}
\affiliation{Beijing Key Lab of Nanophotonics \& Ultrafine Optoelectronic Systems, School of Physics, Beijing Institute of Technology, Beijing, 100081, China}

\author{Run-Wu Zhang}
\affiliation{Centre for Quantum Physics, Key Laboratory of Advanced Optoelectronic Quantum Architecture and Measurement (MOE), School of Physics, Beijing Institute of Technology, Beijing, 100081, China}
\affiliation{Beijing Key Lab of Nanophotonics \& Ultrafine Optoelectronic Systems, School of Physics, Beijing Institute of Technology, Beijing, 100081, China}

\author{Zhi-Ming Yu}
\email{zhiming\_yu@bit.edu.cn}
\affiliation{Centre for Quantum Physics, Key Laboratory of Advanced Optoelectronic Quantum Architecture and Measurement (MOE), School of Physics, Beijing Institute of Technology, Beijing, 100081, China}
\affiliation{Beijing Key Lab of Nanophotonics \& Ultrafine Optoelectronic Systems, School of Physics, Beijing Institute of Technology, Beijing, 100081, China}
\affiliation{International Center for Quantum Materials, Beijing Institute of Technology, Zhuhai, 519000, China}

\author{Yugui Yao}
\email{ygyao@bit.edu.cn}
\affiliation{Centre for Quantum Physics, Key Laboratory of Advanced Optoelectronic Quantum Architecture and Measurement (MOE), School of Physics, Beijing Institute of Technology, Beijing, 100081, China}
\affiliation{Beijing Key Lab of Nanophotonics \& Ultrafine Optoelectronic Systems, School of Physics, Beijing Institute of Technology, Beijing, 100081, China}
\affiliation{International Center for Quantum Materials, Beijing Institute of Technology, Zhuhai, 519000, China}

\begin{abstract}
Despite the rapid progress in the study of  planar Hall effect (PHE) in recent years, all the previous works only showed that the PHE is connected  to local geometric quantities, such as Berry curvature. 
Here, \emph{for the first time}, we point out that the PHE in magnetic Weyl semimetals is  directly related to a global quantity, namely, the Chern number of the Weyl point.
This leads to a  remarkable consequence that the PHE observation predicted here is robust against many system details,  including the Fermi energy.
The main  difference between non-magnetic and magnetic Weyl points is that the latter breaks time-reversal symmetry ${\cal{T}}$, thus generally possessing an energy tilt.
Via semiclassical Boltzmann theory, we investigate the PHE in  generic magnetic Weyl models with energy tilt and arbitrary Chern number. We find that by aligning the magnetic and electric fields in the same direction, the  trace of the PHE conductivity contributed from Berry curvature  and orbital moment is  proportional to the Chern number and the  energy tilt of the Weyl points, resulting in previously undiscovered quantized PHE plateau by varying Fermi energy.
We further confirm the existence of PHE plateaus in a more realistic lattice model without  ${\cal{T}}$ symmetry.
By proposing a new quantized physical quantity, our work not only   provides a new tool  for extracting  the topological character of the Weyl points but also suggests  that the interplay between topology and magnetism can give rise to intriguing  physics.
\end{abstract}

\maketitle
\textcolor{blue}{\textit{Introduction.}}---
The band topology, referring to the topological properties of the electronic band structure of materials, have attracted considerable attention in the field of condensed matter physics  \cite{RevModPhys.82.1959,RevModPhys.88.021004}.
Generally, one  can utilize  the geometric quantities, which are locally defined at each momentum point in the Brillouin zone (BZ), to describe it. 
Some common geometric quantities include Berry curvature  and its dipole  \cite{RevModPhys.82.1959,PhysRevLett.92.037204,PhysRevLett.115.216806,xu2018electrically,PhysRevB.97.041101,PhysRevLett.130.016301}, Berry-connection polarizability and its dipole  \cite{PhysRevLett.127.277202,PhysRevLett.127.277201,PhysRevB.105.045118,PhysRevB.107.205120,PhysRevLett.112.166601,PhysRevLett.129.086602}, and quantum metric \cite{PhysRevB.94.134423,rhim2020quantum,ahn2022riemannian,PhysRevB.108.L201405,PhysRevB.109.115121}.
However, the essential  aspect of topological states is that they host (at least) a global quantity, i.e. topological charge \cite{RevModPhys.88.035005,xie2021higher,bradlyn2017topological,po2017symmetry,PhysRevX.7.041069}, which also is used to characterize and distinguish topological states.
For example, the integral of the Berry curvature over a 2D closed surface in momentum space may result in Chern number \cite{vanderbilt2018berry}.
More importantly, in stark contrast to local geometric quantities and general physical observations, global topological quantities are integers and are robust to many material details  \cite{bernevig2013topological,shun2018topological}.

Recently, various Hall effects, including nonlinear Hall effects \cite{du2021nonlinear,PhysRevLett.127.277202,PhysRevLett.127.277201,PhysRevB.100.165422,PhysRevB.100.195117,lai2021third,PhysRevB.105.045118,PhysRevLett.131.056401,PhysRevB.107.205120,PhysRevB.107.245141,wang2023quantum,gao2023quantum,
kaplan2023general,PhysRevB.107.075411,PhysRevLett.131.076601,PhysRevLett.130.166302,PhysRevB.107.155109,PhysRevLett.132.106601,
PhysRevLett.132.096302,fang2023quantum,huang2023scaling,PhysRevLett.132.026301} and many kinds of planar Hall effect (PHE) \cite{doi:10.1063/1.364592,PhysRevLett.123.016801,PhysRevB.102.241105,PhysRevResearch.3.L012006,PhysRevLett.130.166702,PhysRevLett.130.126303,wang2023fieldinduced,
PhysRevB.108.L241104,PhysRevB.108.075155,zheng2024layer,PhysRevB.109.075419,PhysRevLett.132.056301}, have been proposed and demonstrated to have a close relationship with geometric quantities. Since the geometric quantities are significantly enhanced around band degeneracies, these unconventional Hall effects are widely investigated in topological semimetals \cite{PHE2,PHE13,PHE5,PHE6,PHE7,PHE8,PHE10,PHE4,PHE11,PHE12}.
As aforementioned, the existence of a global quantity rather than local geometric quantities is the most remarkable  feature of topological states.
Unfortunately, the  PHEs have not yet been reported to be associated with a  global quantity.

\begin{figure}
	\begin{centering}
		\includegraphics[width=\linewidth]{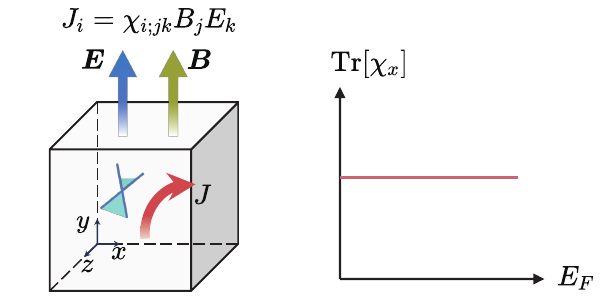}
		\par\end{centering}
	\caption{Sketch of PHE in topological Weyl semimetals with  magnetic and electric fields aligned in the same direction. In this case, the trace of the PHE conductivity ${\rm Tr}[\chi_i] = \sum_{j}\chi_{i; jj}$ is in direct proportion to the Chern number of the WP, and is robust against many system details, leading to a  flat plateau at low energy.	
		\label{fig:fig0}}
\end{figure}

In this work, we show  that the  PHE  in magnetic Weyl semimetals is directly linked to  the topological charge (Chern number) of  the Weyl point.
In PHE,  the magnetic field, the driving electric field, and the transverse Hall current are in the same plane \cite{PHE13,PhysRevB.108.085120}, as illustrated in Fig. \ref{fig:fig0}, which is completely different from the setup in ordinary Hall effect.
Initially, PHE was attributed to system anisotropy \cite{doi:10.1063/1.364592}.
Subsequent findings unveiled that the Berry curvature  also affects PHE, leading to  a surge in research on PHE in topological Weyl semimetals \cite{PhysRevLett.119.176804,PhysRevB.99.115121,PhysRevB.105.205207,PhysRevB.99.165146,PhysRevB.98.205139,PhysRevB.105.205126,
PhysRevB.102.205107,PhysRevB.107.L081110}.
Recently, Li \ea~\cite{PhysRevB.108.085120} found that the PHE resulting from  orbital moment is comparable with that from Berry curvature.
However, a common observation is that the PHE mechanisms in these studies are independent of any global quantity.

Here, based on semiclassical Boltzmann theory, we investigate PHE in the charge-$n$ (with $n$ an integer) Weyl points (C-$n$ WPs) with energy tilt, which is a common feature for  magnetic WPs.
For magnetic WPs, the leading order of PHE generally is linear due to the breaking of time reversal symmetry (${\cal{T}}$) \cite{PhysRevB.108.085120}.
The linear PHE conductivity contains six  terms, with one term contributed by  Lorentz force and  the remaining terms originating from  geometric quantities, i.e. Berry curvature and orbital moment.
Remarkably, by aligning the magnetic  and electric fields in the same direction (see Fig. \ref{fig:fig0}),
we demonstrate that the trace of the  PHE conductivity contributed by Berry curvature  and orbital moment  (Tr[$\chi_i$] with $i=x,y,z$) is quantized for all the C-$n$ WPs resulting from the global topology of the WP, and then is robust against  many material details.
The analytical derivation of  Tr[$\chi_i$] shows that  the  Tr[$\chi_i$] is  fully determined by the Chern number and the energy tilt of the WPs.
This feature becomes increasingly more precise when the Fermi energy approaches the WPs.
By unveiling a novel magnetotransport phenomenon directly related to the global quantities of  WP in magnetic Weyl semimetals, our work deepens the  understanding on the PHE, and provides a promising method to detect the topological charge and the energy tilt of the WPs.


\textcolor{blue}{\textit{Linear PHE conductivity.}}---
Generally, the PHE conductivity under weak fields can be obtained from  semiclassical equations of motion  \cite{RevModPhys.82.1959} and the Boltzmann transport equation  \cite{ashcroftsolid}. 
Here, we focus on the  setup that  the driving electric field $\bm E$ and the magnetic field $\bm B$ are aligned in the same direction, \ie ${\bm E}\|{\bm B}$ (see Fig. \ref{fig:fig0}).
For the three-dimensional magnetic systems, the leading terms of the PHE  that have topological origin are  linearly dependent on both $\bm E$ and $\bm B$, expressed as  \cite{PhysRevB.108.085120}
\begin{align}\label{J_i}
	J_i = \chi_{i;jj} B_j E_j,
\end{align}
where  $\bm J$ is the current density and $\chi_{i;jj}$ denotes the PHE conductivity.
The PHE conductivity $\chi_{i;jk}$  is a third-order tensor, and the  $\chi_{i;jj}$ in Eq. (\ref{J_i}) denotes a diagonal element of the matrix $\chi_i$.
$\chi_{i;jj}$ can be further decomposed into the following five parts \cite{PhysRevB.108.085120}:
\begin{align}
	\chi_{i;jj}  =  \chi_{i; jj}^{(1)} + \chi_{i; jj}^{(2)} + \chi_{i; jj}^{(3)} + \chi_{i; jj}^{(4)} + \chi_{i; jj}^{(5)} ,
\end{align}
with
\begin{alignat}{2}\label{LMC}
	\chi_{i; jj}^{(1)} &= &&-e^{3}\tau \int [d \bm{k} ] v_i v_j \Omega_j \partial_{\varepsilon}f_k^0, \notag \\
	\chi_{i; jj}^{(2)} &= &&e^{3}\tau \int [d \bm{k} ] v_j \left( \bm {v} \cdot \bm \Omega \right) \partial_{\varepsilon}f_k^0, \notag \\
	\chi_{i; jj}^{(3)} &= &&e^{3}\tau \int [d \bm{k} ] v_i \left( \bm {v} \cdot \bm \Omega \right) \partial_{\varepsilon}f_k^0, \notag \\
	\chi_{i; jj}^{(4)} &= &&-e^2\tau \int [d \bm{k} ] v_i \partial_{k_j} m_j  \partial_{\varepsilon}f_k^0, \notag \\
	\chi_{i; jj}^{(5)} &= &&-e^2\tau \int [d \bm{k} ] v_j \partial_{k_i} m_j  \partial_{\varepsilon}f_k^0 .
\end{alignat}
Here, $\int [d \bm{k} ]\equiv -\frac{1}{(2\pi)^3\hbar} \int d^3 k $, $\varepsilon$ is the band energy, $\bm{v} = \partial_{\bm k} \varepsilon/\hbar $ denotes the velocity, $\bm \Omega$  and $\bm m$ respectively are the Berry curvature and the orbital moment,   $\tau$ is the relaxation time, and $f_k^0$ is the Fermi-Dirac distribution. The electron charge is taken as $-e$ (\ie, $e > 0$).

An analysis of Eq. (\ref{LMC}) reveals two crucial features of the topological PHE conductivities $\chi_{i;jj}$.
First, all the five  PHE conductivities linearly depend on $\tau$.
Thus,  they can be easily extracted in experiments \cite{RN2811,RN2812}, as the contributions from intrinsic  Hall effect and Lorentz force are  even functions of $\tau$ \cite{PhysRevB.108.085120}.
This linear  dependence also  indicates that  the topological  PHE conductivities  only appear in the magnetic systems, as ${\cal{T}}$ reverses ${\bm J}$, ${\bm B}$ and $\tau$, but keeps ${\bm E}$ invariant.

Second, one observes that the trace of the matrix $\chi_{i}^{(1)}$ equals to $-\chi_{i; jj}^{(3)}$, \ie ${\rm Tr}[\chi_i^{(1)}] = \sum_{j}\chi_{i; jj}^{(1)} = -\chi_{i; jj}^{(3)}$, and ${\rm Tr}[\chi_i^{(3)}] = 3\chi_{i; jj}^{(3)}$. 
Thus, the trace of  $\chi_{i}$ is 
\begin{align}
	{\rm Tr}[\chi_i] \equiv \sum_{j}\chi_{i; jj} = {\rm Tr}[\chi_{i}^{(2)} + \frac{2}{3}\chi_{i}^{(3)} + \chi_{i}^{(4)} + \chi_{i}^{(5)}],
\end{align}
in which the contribution from Berry curvature and orbital moment are ${\rm Tr}[\chi_{i}^{\Omega}]={\rm Tr}[\chi_{i}^{(2)} + \frac{2}{3}\chi_{i}^{(3)}]$ and  ${\rm Tr}[\chi_{i}^{m}]={\rm Tr}[\chi_{i}^{(4)} + \chi_{i}^{(5)}]$, respectively.

Third, the topological PHE conductivities $\chi_{i;jj}$ are determined by the properties of the Fermi surface, due to the presence of $\partial_{\varepsilon}f_k^0$.
It has been demonstrated that the geometry quantities at the Fermi surface have a significant influence  on the PHE conductivities \cite{PhysRevB.108.085120,PhysRevLett.119.176804,PhysRevB.99.115121}.
However, the  Fermi surface may also possess  global quantities.
The influence of the global quantities on the PHE conductivities has not  been investigated.
Since the most famous  global quantity that may be hosted by a 2D Fermi surface  is the Chern  number (${\cal{C}}$), here  we  focus on the topological PHE conductivities $\chi_{i;jj}$ of  generic magnetic WPs with different Chern  numbers.

\textcolor{blue}{\textit{Linear WP  and scaling analysis.}}---
Before studying a generic Weyl model with arbitrary Chern number, it is instructive to  begin with a  conventional (linear) WP with $|{\cal{C}}|=1$, which has a linear dispersion along all the directions in momentum space  \cite{YUScienceBulletin}.
Due to the absence of ${\cal{T}}$ symmetry, the  magnetic  WP generally has an energy tilt.
For linear WP, the leading term for the energy tilt is also  linear.
Thus, a   general Hamiltonian for the magnetic linear WP can be written as  \cite{YUScienceBulletin}
\begin{align}\label{model1}
	\mathcal{H}_1 = w_x k_x + v_1 \bm k \cdot \bm\sigma,
\end{align}
with  $w_x$ and $v_1$ the   real parameters respectively representing the energy tilt and Fermi velocity of the WP, and $\sigma$ being the Pauli matrix.
Here, without loss of generality, we have set the  energy tilt to be along $k_x$ direction. We also  assume $|w_x|<|v_1|$ to guarantee the Fermi surface of the WP close  \cite{soluyanov2015type,PhysRevLett.117.077202}.
It is easy to know that this model  (\ref{model1}) exhibits a linear dispersion  and exhibits a Chern number ${\cal{C}}={\rm Sign} (v_1)$.

\begin{figure}
	\begin{centering}
		\includegraphics[width=\linewidth]{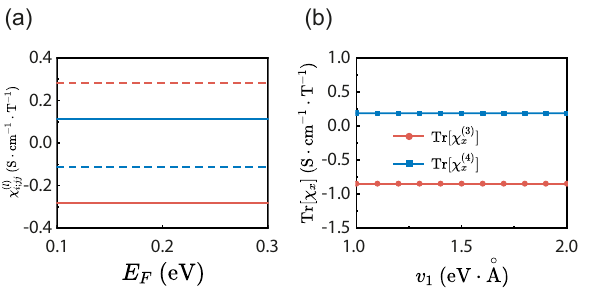}
		\par\end{centering}
	\caption{
Variation of PHE conductivity 
with (a) Fermi level $E_F$ and (b) Fermi velocity  $v_1$. In panel (a), the red and blue lines correspond to the $\chi_{x;yy}^{(3)}$ and $\chi_{x;xx}^{(4)}$, respectively. Solid lines represent results for $w_x > 0$ and dashed lines for $w_x < 0$.
We set $|w_x| = 0.2 \ \text{eV}\cdot\text\AA$, and $v_1 = 1.0 \ \text{eV}\cdot\text\AA$ in (a), and set $w_x =0.2 \ \text{eV}\cdot\text\AA$ and $E_F= 200 ~\rm{meV}$ in (b). We also set   $\tau = 0.1~\rm{ps}$ in all calculations.
		\label{fig:fig1}}
\end{figure}


A striking feature of the system that exhibits same  order of dispersion along all the directions is that its various important properties can be obtained by applying a simple scaling analysis \cite{Caojin}.
For this model (\ref{model1}), we consider a scaling transformation in momentum and energy: $\bm k \rightarrow \lambda \bm k$ and $E_F \rightarrow \lambda E_F $, where  $\lambda$ is a real number.
Due to the linear feature of Eq. (\ref{model1}), one has ${\cal{H}}_1(\lambda \bm k)=\lambda{\cal{H}}_1( \bm k)$ and $\varepsilon(\lambda \bm k)=\lambda \varepsilon(\bm k)$.
Note that the velocity and  the eigenstates of the system are invariant under the scaling, \ie $v(\lambda \bm k)=v(\bm k)$ and $|u_{n,\lambda \bm k}\rangle=|u_{n,\bm k}\rangle$.
Then, the Berry curvature and orbital moment transform as  ${\bm{\Omega}_{\lambda \bm{k}}}=\lambda^{-2} {\bm{\Omega}_{ \bm{k}}}$ and ${\bm m}_{\lambda \bm{k}} = \lambda^{-1} {\bm m}_{\bm{k}}$, respectively.

\begin{figure}
	\begin{centering}
		\includegraphics[width=\linewidth]{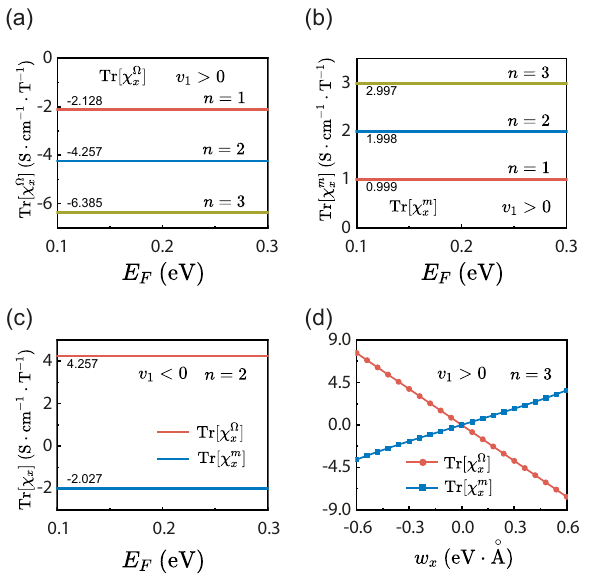}
		\par\end{centering}
	\caption{
Variation of PHE conductivity (a) $\mathrm{Tr}[\chi^{\Omega}_x]$ and (b) $\mathrm{Tr}[\chi^m_x]$ with Fermi level for $n = 1, 2, 3$ when $v_1 > 0$. The corresponding value of $\eta$ and $\beta$ in Table \ref{tab} are $\eta = 0.474~\rm{S \cdot \AA \cdot cm^{-1}\cdot eV^{-1} \cdot T^{-1}}$ and $\beta = 0.112$. 
(c) Variation of $\mathrm{Tr}[\chi^{\Omega}_x]$ and  $\mathrm{Tr}[\chi^m_x]$ with Fermi level for $n = 2$ when $v_1 < 0$. The corresponding $\beta = 0.143$.
(d)  Variation of  $\mathrm{Tr}[\chi^{\Omega}_x]$ and  $\mathrm{Tr}[\chi^m_x]$ with energy tilt $w_x$ for $n = 3$ when $v_1 > 0$.
We set $w_x = 0.5 \ \text{eV}\cdot\text\AA$ in (a-c), $v_1=1.0 \ \text{eV}\cdot\text\AA$,  $c_+ = 0.3 \ \text{eV}\cdot\text\AA^n$, and $c_- = 1.2 \ \text{eV}\cdot\text\AA^n$ in (a), (b) and (d); $v_1=-0.9 \ \text{eV}\cdot\text\AA$, $c_+ = 0.4 \ \text{eV}\cdot\text\AA^n$ and $c_- = 1.0 \ \text{eV}\cdot\text\AA^n$ in (c).
We also set $E_F = 200~\text{meV}$ in (d) and $\tau = 0.1~\rm{ps}$ in all calculations.
		\label{fig:fig2}}
\end{figure}

As a consequence, for all the five PHE conductivities in  Eq. (\ref{LMC}), we have
\begin{align}
	\chi_{i;jj}^{(l)}(\lambda E_F) =\chi_{i;jj}^{(l)}(E_F),
\end{align}
indicating that the $\chi_{i;jj}^{(l)}$ ($l = 1,\cdots, 5$) is a quantity that surprisingly has no dependence on Fermi energy, leading to a PHE plateau in energy space, as shown in Fig. \ref{fig:fig1}(a).
Further  numerical calculations show that the trace of the PHE conductivity ${\rm Tr}[\chi_{i}^{(l)}]$ is also robust to  the absolute value of $v_1$, and is only dependent on its sign  [see Fig. \ref{fig:fig1}(b)], which  is reminiscent of the Chern number of the WP  ${\cal{C}}={\rm Sign} (v_1)$.
Notice that in the field of condensed matter physics, it is quite rare for an \emph{observable} to have such a robust feature.
All these results indicate  that the PHE conductivity ${\rm Tr}[\chi_{i}^{(l)}]$ in WP should be  more nontrivial than previously understood.
Besides, from  symmetry analysis, one knows that  the value of the PHE plateau would  be reversed when the sign of $w_x$ is changed, as both $w_x$ and $\chi_{i,jj}^{(l)}$ are anti-symmetric  under  ${\cal{T}}$ symmetry.
This is confirmed by our numerical calculations, as shown in Fig. \ref{fig:fig1}(a).

\textcolor{blue}{\textit{Beyond  linear model.}}---
We then discuss the trace of PHE conductivity ${\rm Tr}[\chi_{i}]$  in a generic magnetic WP with arbitrary Chern number, for which the Hamiltonian reads   \cite{PhysRevB.105.085117,PhysRevB.105.104426},
\begin{align}\label{model2}
	\mathcal{H}_2=w_x k_x + \left[\begin{array}{cc}
		v_1 k_x & c_{+} k_{+}^n+c_{-} k_{-}^n \\
		c_{+} k_{-}^n+c_{-} k_{+}^n & -v_1 k_x
	\end{array}\right],
\end{align}
with $k_\pm = k_y \pm ik_z$, $n$ an integer (which can be 1, 2, and 3 in crystals), and $c_\pm$ a real parameter.
The Chern number of this WP is ${\cal{C}}=n\times {\rm Sign}(v_1)$  for $|c_+|>|c_-|$ and is ${\cal{C}}=-n\times {\rm Sign}(v_1)$ for $|c_+|<|c_-|$.
Again, we require the WP to be  a type-I point with  $|w_x|<|v_1|$.
For $n\neq1$, the WP has a linear dispersion along $k_x$, but a nonlinear dispersion along other directions.
Thus, it can not be investigated by scaling analysis.

The numerical results of  ${\rm Tr}[\chi_{i}^{\Omega(m)}]$ based on Hamiltonian $\mathcal{H}_2$ (\ref{model2}) are presented in  Fig. \ref{fig:fig2}.
A remarkable observation is that while $\chi_{i,jj}^{\Omega(m)}$  is no longer independent of Fermi energy, due to the loss of linear dispersion, the trace, i.e. ${\rm Tr}[\chi_{i}^{\Omega(m)}]$ remain robust against $E_F$, as shown in Fig. \ref{fig:fig2}(a-c).
More than that, the   plateau of ${\rm Tr}[\chi_{i}^{\Omega(m)}]$  is  completely not effected by the nonlinear dispersion and the band anisotropy induced by the parameters $c_\pm$ (see Fig. \ref{fig:fig2}).
Again, we  find that the PHE plateau   is sensitive to the value of $n$ and the sign of $v_1$, \ie the Chern number of the WP.
Moreover, we find that ${\rm Tr}[\chi_{i}^{\Omega}]$ linearly depend  on the energy tilt $w_x$, while ${\rm Tr}[\chi_{i}^{m}]$ shows a linear dependence on $w_x$ when $w_x$ is small, but increasingly  deviates it when $w_x$ increases, as shown in Fig. \ref{fig:fig2}(d).
All these results strongly suggest that the trace of the  PHE conductivity ${\rm Tr}[\chi_{i}^{\Omega(m)}]$ has a deeper physical origin beyond local geometric quantities.

\global\long\def\arraystretch{1.4}%
\begin{table}[t]
	\caption{The PHE conductivity ${\rm Tr}[\chi_{i}^{(l)}]$ of a generic WP model (\ref{model2}) where the energy tilt is along $k_x$ direction. Here, $\mathcal C$ is the Chern number of the WP,  $\eta  = e^3 \tau/(12\pi^2\hbar^3)$, and $\beta  = [-2\alpha^3 + 3\alpha + 3(\alpha^2 -1)\operatorname{arctanh} \alpha]/\alpha^3$. $w_x$ and $\alpha=|w_x/v_1| < 1$ denote the absolute and relative energy tilt, respectively. }
	\begin{ruledtabular}
		\begin{tabular}{ccccccc}
			\multicolumn{1}{c}{}&\multicolumn{3}{c}{Berry curvature} & \multicolumn{2}{c}{orbital moment} \\
			\hline
			 & $l$ = 1 & $l$ = 2 & $l$ = 3 & $l$ = 4 & $l$ = 5 \\
			${\rm Tr}[\chi_{x}^{(l)}]/\eta $ & $-3 w_x\mathcal{C}$ & $3 w_x\mathcal{C}$ & $9 w_x\mathcal{C}$ &  $(-2-\beta)w_x\mathcal{C}$ & $(-2-\beta)w_x\mathcal{C}$  \\
			${\rm Tr}[\chi_{y}^{(l)}]/\eta$ & 0 & $3 w_x\mathcal{C}$ & 0 &  0 & 0 \\
			${\rm Tr}[\chi_{z}^{(l)}]/\eta$ & 0 & $3 w_x\mathcal{C}$ & 0 &  0 & 0 \\
		\end{tabular}
	\end{ruledtabular}
	\label{tab}
\end{table}

\textcolor{blue}{\textit{Analytical expression of ${\rm Tr}[\chi_{i}]$.}}---Next, we  provide a rigorous explanation for the novel and intriguing  results obtained above.
Due to the presence $\partial_{\varepsilon}f_k^0$,  at zero temperature the integral in Eq. (\ref{LMC}) actually is performed on the Fermi surface.
Since ${\rm Tr}[\chi_i^{(1)}] = -\chi_{i; jj}^{(3)}=-{\rm Tr}[\chi_{i}^{(3)}]/3$, we  start our analysis from ${\rm Tr}[\chi_{i}^{(3)}]$, which can be rewritten as
\begin{equation}\label{chi}
	{\rm Tr}[\chi_{i}^{(3)}]= 3\frac{e^3\tau}{(2\pi)^3\hbar^2}\oint_{\varepsilon = E_F} v_i\Omega_\perp dS,
\end{equation}
where $S$ denotes the Fermi surface, the subscript $\perp$  represents the normal component of the corresponding quantities with respect to $S$, and the equations $\bm{v_k} \cdot \bm {\Omega_k} = v_\perp\Omega_\perp$ and  $v_\perp = \partial_{k_\perp}\varepsilon/\hbar$ are used.
For model (\ref{model2}), its dispersion is $\varepsilon=w_x k_x +\varepsilon_0$ with $\varepsilon_0$ denotes the dispersion without energy tilt.
Then, one has $v_x =w_x/\hbar + v_{x,0}$ with $v_{x,0}=\partial_{k_x}\varepsilon_0/\hbar$.
Interestingly, the $x$-component of ${\rm Tr}[\chi^{(3)}]$ becomes [see the Supplementary Material \cite{suppmat} for details]
\begin{align}\label{chi2}
	{\rm Tr}[\chi_{x}^{(3)}] &= \frac{3e^3\tau}{(2\pi)^3\hbar^3}(w_x\oint \Omega_\perp dS+\oint  \hbar v_{x,0}\Omega_\perp dS) \notag \\
&=9\eta   w_x \mathcal C,
\end{align}
with $\eta  = \frac{e^3 \tau}{12\pi^2\hbar^3}$ and $\mathcal C$ the Chern number of the 2D Fermi surface, which is exactly the Chern number of the WP. 
Since  the energy tilt along $k_{y(z)}$ is zero, one has   ${\rm Tr}[\chi_{y}^{(3)}]={\rm Tr}[\chi_{z}^{(3)}]=0$ \cite{suppmat}.
Besides, from Eq. \ref{LMC}, we have ${\rm Tr}[\chi^{(2)}_i] = ({\rm Tr}[\chi_{x}^{(3)}]+{\rm Tr}[\chi_{y}^{(3)}]+{\rm Tr}[\chi_{z}^{(3)}])/3$, indicating that  ${\rm Tr}[\chi_{x/y/z}^{(2)}]=3\eta  w_x\mathcal C$.
Therefore,  the PHE conductivity contributed by Berry curvature  is
\begin{eqnarray}\label{m1}
	{\rm Tr}[\chi_{x}^{\Omega}] & = & 9\eta  w_x \mathcal C, \\
    {\rm Tr}[\chi_{y}^{\Omega}] & = & {\rm Tr}[\chi_{z}^{\Omega}]=3\eta  w_x \mathcal C.
\end{eqnarray}
A more complicated calculation demonstrates that ${\rm Tr}[\chi_{i}^{(4)}]={\rm Tr}[\chi_{i}^{(5)}]$, and   the  PHE conductivity contributed by orbital moment  is \cite{suppmat}
\begin{eqnarray}
	{\rm Tr}[\chi_{x}^{m}] &= & -(4 + 2\beta) \eta  w_x \mathcal C,  \\
 {\rm Tr}[\chi_{y}^{m}] &= & {\rm Tr}[\chi_{z}^{m}]=0, \label{m2}
\end{eqnarray}
with $\beta  = [-2\alpha^3 + 3\alpha + 3(\alpha^2 -1)\operatorname{arctanh} \alpha]/\alpha^3$ and $\alpha=|w_x/v_1| < 1$ denoting the  relative energy tilt.
Obviously, $\beta$ represents the deviation of ${\rm Tr}[\chi^{(m)}_{x}]$ from a simple linear relationship with $w_x{\cal{C}}$.  It  rapidly approaches zero as $w_x$ decreases, which can be inferred from the alternative  expression of $\beta$:    $\beta  = \sum_{n = 1}^{\infty} \frac{6 \alpha^{2n}}{4n^2+8n+3}$.
The Eqs. (\ref{m1}-\ref{m2}) are the main results of this work, and  is summarized in Table \ref{tab}.
These results directly show that the PHE is highly nontrivial, and originates from the global topological quantities of the Fermi surface.
Consequently, in contrast to various Hall effects, the PHE conductivity in magnetic Weyl semimetals are robust against many kinds of small perturbations, as shown in   Fig. \ref{fig:fig2}.

\begin{figure}[b]
	\begin{centering}
		\includegraphics[width=\linewidth]{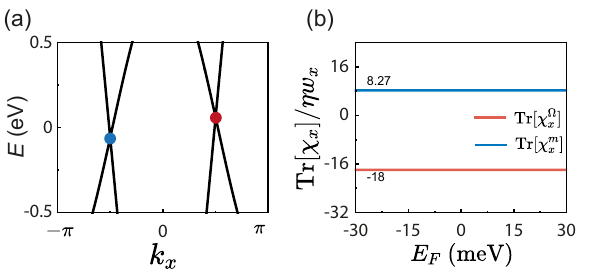}
		\par\end{centering}
	\caption{
		(a) Band structure of the Weyl model (\ref{model3}). (b) Variation of the PHE conductivity with Fermi energy where the corresponding $\beta = 0.069$. 
We set   $\tau = 0.1~\rm{ps}$, $m = t = 1.5 \ \text{eV}\cdot\text\AA$, $w_x = 0.6 \ \text{eV}\cdot\text\AA$,  $\Delta = 0.06 \ \text{eV}\cdot\text\AA$ and $k_0 = \pi/2 \ \text\AA^{-1}$.
		\label{fig:latticeModel}}
\end{figure}

\textcolor{blue}{\textit{Lattice model.}}---Our results so far are obtained on the  effective model around a WP.  We point out that the PHE plateau can persist  in a  realistic lattice model, and is almost precise when the Fermi energy is around WPs. Consider a minimal lattice models for topological magnetic Weyl semimetal, expressed as \cite{PhysRevB.95.075133}
\begin{align}\label{model3}
	\mathcal{H}_3= & \left[ m\left(\cos k_y + \cos k_z -2 \right) - t\left( \cos k_x -\cos k_0 \right) \right] \sigma_x  \notag \\
	& - t \sin k_y \sigma_y - t \sin k_z \sigma_z + w_x \left( \cos k_0-\cos k_x \right) \notag \\
	& + \Delta \sin k_x,
\end{align}
where  $m$, $t$  and $k_0$ are model parameters, and $w_x$ denotes the energy tilt.
The band structure of the Weyl model (\ref{model3}) is plotted in Fig. \ref{fig:latticeModel}(a), in which two  tilted linear   WPs with $|{\cal C}=1|$ at $(\pm k_0,0,0)$ position can be observed when  $w_x$ is finite.
Since the two WPs have both opposite energy tilt and Chern number, one can expect that the total PHE conductivity ${\rm Tr}[\chi_{i}] (\propto w_x{\cal C})$ from the two WPs  would not be canceled but is added.

The numerically calculated PHE conductivities  of model (\ref{model3}) as a function of $E_F$ are plotted in Fig. \ref{fig:latticeModel}(b).
These conductivities continue to display a flat plateau and preserve their mutual proportional relationships, as revealed in Eqs. (\ref{m1}-\ref{m2}) and Table \ref{tab}.
Moreover, the plateau values have now doubled due to the presence of two WPs.

\textcolor{blue}{\textit{Conclusions.}}---
In conclusion, we demonstrate that the PHE conductivity can be directly related to the topological charge of WPs, and then features quantized plateau by varying many material parameters. 
Our work significantly advances the understanding of magnetotransport in magnetic Weyl semimetals, highlighting the crucial role of Chern number and energy tilt of the WPs,  and providing an attractive prospect for future explorations on quantized Hall transport phenomena in topological materials.

\acknowledgments
This work was supported by the National Key R\&D Program of China (Grant No. 2020YFA0308800), the NSF of China (Grants Nos. 12004035, 12234003 and 12321004) and the China Postdoctoral Science Foundation (Grants Nos. 2021TQ0043 and 2021M700437).

\bibliography{myRef.bib}

\end{document}


\title{Supplemental Material for “ Planar Hall Plateau in Magnetic Weyl Semimetals ”}

\author{Lei Li}
\thanks{These authors contributed equally to this work}
\affiliation{Centre for Quantum Physics, Key Laboratory of Advanced Optoelectronic Quantum Architecture and Measurement (MOE), School of Physics, Beijing Institute of Technology, Beijing, 100081, China}
\affiliation{Beijing Key Lab of Nanophotonics \& Ultrafine Optoelectronic Systems, School of Physics, Beijing Institute of Technology, Beijing, 100081, China}

\author{Chaoxi Cui}
\thanks{These authors contributed equally to this work}
\affiliation{Centre for Quantum Physics, Key Laboratory of Advanced Optoelectronic Quantum Architecture and Measurement (MOE), School of Physics, Beijing Institute of Technology, Beijing, 100081, China}
\affiliation{Beijing Key Lab of Nanophotonics \& Ultrafine Optoelectronic Systems, School of Physics, Beijing Institute of Technology, Beijing, 100081, China}

\author{Run-Wu Zhang}
\affiliation{Centre for Quantum Physics, Key Laboratory of Advanced Optoelectronic Quantum Architecture and Measurement (MOE), School of Physics, Beijing Institute of Technology, Beijing, 100081, China}
\affiliation{Beijing Key Lab of Nanophotonics \& Ultrafine Optoelectronic Systems, School of Physics, Beijing Institute of Technology, Beijing, 100081, China}

\author{Zhi-Ming Yu}
\email{zhiming\_yu@bit.edu.cn}
\affiliation{Centre for Quantum Physics, Key Laboratory of Advanced Optoelectronic Quantum Architecture and Measurement (MOE), School of Physics, Beijing Institute of Technology, Beijing, 100081, China}
\affiliation{Beijing Key Lab of Nanophotonics \& Ultrafine Optoelectronic Systems, School of Physics, Beijing Institute of Technology, Beijing, 100081, China}

\author{Yugui Yao}
\email{ygyao@bit.edu.cn}
\affiliation{Centre for Quantum Physics, Key Laboratory of Advanced Optoelectronic Quantum Architecture and Measurement (MOE), School of Physics, Beijing Institute of Technology, Beijing, 100081, China}
\affiliation{Beijing Key Lab of Nanophotonics \& Ultrafine Optoelectronic Systems, School of Physics, Beijing Institute of Technology, Beijing, 100081, China}

\maketitle
\section{Derivation of the PHE conductivity}
\subsection{Linear model}
We begin with the   linear WP model described by Eq. (5) in the main text, as it is simple and its PHE conductivity can be  obtained directly.
The  energy dispersion of this linear WP  is $\varepsilon_{\bm k} = w_xk_x\pm v_1 k$ with $k=\sqrt{k_x^2+k_y^2+k_z^2}$.
For the conduction band,  the Berry curvature and the orbital moment are $\bm{\Omega_k} = -{\rm Sign}(v_1) \bm k /(2k^3)$ and  $\bm m = -{\rm Sign}(v_1) e v_1 \bm k/(2k^2\hbar)$, respectively.  Specifically, we set $v_1 > 0$. Then  we have  $\bm{\Omega_k} = - \bm k /(2k^3)$, $\bm m = -e v_1 \bm k/(2k^2\hbar)$ and the Chern number ${\cal{C}}=-1$ for conduction band.
Since  the energy tilt of this WP is  along the $k_x$ direction, it is convenient to calculate the PHE conductivity in   spherical coordinates, expressed as:
\begin{equation}
	k_x = k \cos\theta, \ \ \ k_y = k \sin\theta \cos\phi,  \ \ \ 	k_z= k \sin\theta \sin\phi.
\end{equation}
In the zero-temperature limit, we have $\partial_{\varepsilon}f_k^0 = -\delta(\varepsilon - E_F)$, which implies that the integrals for all the PHE conductivities  here are  performed on the Fermi surface.
Then, for $\chi^{(3)}_{x;jj}$, one has
\begin{align}
	\chi^{(3)}_{x;jj} &= -\frac{e^3 \tau}{\hbar} \int \frac{d^3 k }{(2\pi)^3} v_x \left( \boldsymbol{\Omega}_{\bm{k}} \cdot \bm {v}_{\bm{k}} \right) \frac{\partial f_0}{\partial \varepsilon}  \notag \\
	&= \frac{e^3 \tau}{\hbar^3} \int \frac{d^3 k }{(2\pi)^3} \left( -\frac{k_x v_1^2}{2 k^3} - \frac{v_1 w_x}{2 k^2} - \frac{k_x^2 v_1 w_x}{2 k^4} - \frac{k_x w_x^2}{2 k^3} \right) \delta(v_1 k + k_xw_x - E_F)  \notag \notag \\
&= \frac{e^3 \tau}{\hbar^3} \frac{1}{(2\pi)^3}\int d\phi d\theta  \sin\theta  \frac{-\cos \theta v_1^2 - v_1 w_x - v_1 w_x \cos^2 \theta - w_x^2 \cos \theta}{2(v_1+ w_x \cos \theta)} \int d k (v_1+ w_x \cos \theta)\delta[k (v_1+ w_x \cos \theta)   - E_F]  \notag \notag \\
	&= \frac{e^3 \tau}{\hbar^3} \frac{1 }{(2\pi)^3} \int_{0}^{\pi} d\theta \left[ -\pi \left( w_x + v_1\cos\theta\right) \sin\theta \right] \notag \\
	&= -\frac{e^3 \tau w_x}{4\pi^2\hbar^3} \notag \\
	&= 3\eta w_x {\cal{C}},
\end{align}
with  $\eta  = \frac{e^3 \tau }{12\pi^2\hbar^3}$. This naturally leads to the results in Table I in main text, namely,
\begin{equation}
	{\rm Tr}[\chi^{(3)}_x]= 9\eta w_x\mathcal C.
\end{equation}
Similarly, for $\chi^{(3)}_{y;jj}$ and $\chi^{(3)}_{z;jj}$, one has
\begin{align}
	\chi^{(3)}_{y;jj} &= -\frac{e^3 \tau}{\hbar} \int \frac{d^3 k }{(2\pi)^3} v_y \left( \boldsymbol{\Omega}_{\bm{k}} \cdot \bm {v}_{\bm{k}} \right) \frac{\partial f_0}{\partial \varepsilon}  \notag \\
	&= \frac{e^3 \tau}{\hbar^3} \int \frac{d^3 k }{(2\pi)^3} \left( -\frac{k_y v_1^2}{2 k^3} - \frac{k_x k_y v_1 w_x}{2 k^4} \right) \delta(v_1 k + k_xw_x - E_F)  \notag \notag \\
	&= \frac{e^3 \tau}{\hbar^3} \frac{1 }{(2\pi)^3} \int_{0}^{\pi} d\theta \int_{0}^{2\pi} d\phi \left( -\frac{1}{2}v_1\sin^2\theta \cos\phi \right) \notag \\
	&= 0,
\end{align}
and
\begin{align}
	\chi^{(3)}_{z;jj} &= -\frac{e^3 \tau}{\hbar} \int \frac{d^3 k }{(2\pi)^3} v_z \left( \boldsymbol{\Omega}_{\bm{k}} \cdot \bm {v}_{\bm{k}} \right) \frac{\partial f_0}{\partial \varepsilon}  \notag \\
	&= \frac{e^3 \tau}{\hbar^3} \int \frac{d^3 k }{(2\pi)^3} \left( -\frac{k_z v_1^2}{2 k^3} - \frac{k_x k_z v_1 w_x}{2 k^4} \right) \delta(v_1 k + k_xw_x - E_F)  \notag \notag \\
	&= \frac{e^3 \tau}{\hbar^3} \frac{1 }{(2\pi)^3} \int_{0}^{\pi} d\theta \int_{0}^{2\pi} d\phi \left( -\frac{1}{2}v_1\sin^2\theta \sin\phi \right) \notag \\
	&= 0,
\end{align}
indicating that
\begin{equation}
	{\rm Tr}[\chi^{(3)}_y] = {\rm Tr}[\chi^{(3)}_z] = 0.
\end{equation}
 According to the Eq. (3) in the main text, we also have
\begin{eqnarray}
  {\rm Tr}[\chi^{(1)}_i] &=& -\chi^{(3)}_{i;jj}, \\
  {\rm Tr}[\chi^{(2)}_i] &=& \chi^{(3)}_{x;jj}+\chi^{(3)}_{y;jj}+\chi^{(3)}_{z;jj}=3\eta w_x\mathcal C.
\end{eqnarray}

For the  terms induced by the orbital moment, \ie ${\rm Tr}[\chi^{(4)}_i]$ and ${\rm Tr}[\chi^{(5)}_i]$, their expressions  are more complex. Specifically, one has
\begin{align}\label{eq:chi4_1}
	{\rm Tr}[\chi^{(4)}_x] &= \sum_j  \chi^{(4)}_{x;jj} \notag \\
	&= \sum_j \frac{e^2 \tau}{\hbar} \int \frac{d^3 k }{(2\pi)^3} v_x \partial_{k_j} m_j \frac{\partial f_0}{\partial \varepsilon} \notag \\
	&= -\frac{e^3 \tau}{\hbar^3} \frac{1 }{(2\pi)^3} \int_{0}^{\pi} d\theta \left[-\frac{\pi v_1(w_x + v_1\cos\theta)\sin\theta}{v_1+w_x\cos\theta}\right] \notag \\
	&= \frac{e^3 \tau}{\hbar^3} \frac{1 }{(2\pi)^3} \int_{0}^{\pi} d\theta \left[\frac{\pi w_x (1+\cos\theta/\alpha)\sin\theta}{1+\alpha\cos\theta} \right] \notag \\
	&= \frac{e^3 \tau}{\hbar^3} \frac{\pi w_x}{(2\pi)^3} \left. \left[-\frac{\cos\theta}{\alpha^2} -\frac{(\alpha^2-1)\ln(1+\alpha\cos\theta)}{\alpha^3} \right] \right |_0^{\pi}  \notag \\
	&= 2\eta  w_x \frac{3\alpha + 3(\alpha^2 -1)\operatorname{arctanh} \alpha}{2\alpha^3}   \notag \\
	&= 2\eta  w_x + \frac{-2\alpha^3 + 3\alpha + 3(\alpha^2 -1)\operatorname{arctanh} \alpha}{\alpha^3}\eta  w_x \notag \\
	&= (2 + \beta) \eta  w_x \notag \\
	&= (-2 - \beta) \eta  w_x \mathcal{C}
\end{align}
where we have defined  $\alpha = |w_x/v_1|$ and $\beta=\frac{-2\alpha^3 + 3\alpha + 3(\alpha^2 -1)\operatorname{arctanh} \alpha}{\alpha^3}$. For type-I WP, one has $\alpha<1$.
To further clarify the physical implications of the correction factor $\beta$,  we  rewrite it as $\beta = \sum_{n = 1}^{\infty} \frac{6 \alpha^{2n}}{4n^2+8n+3}$, which shows the correction is induced by the relative energy tilt $\alpha$, and  rapidly approaches zero when $\alpha$ decreases.

We continue to calculate   ${\rm Tr}[\chi^{(4)}_y]$ and  ${\rm Tr}[\chi^{(4)}_z]$. One has
\begin{align}
	{\rm Tr}[\chi^{(4)}_y] &= \sum_j  \chi^{(4)}_{y;jj} \notag \\
	&= \sum_j \frac{e^2 \tau}{\hbar} \int \frac{d^3 k }{(2\pi)^3} v_y \partial_{k_j} m_j \frac{\partial f_0}{\partial \varepsilon} \notag \\
	&= -\frac{e^3 \tau}{\hbar^3} \frac{1 }{(2\pi)^3} \int_{0}^{\pi} d\theta \int_{0}^{2\pi} d\phi \left[  -\frac{v_1^2 \sin^2 \theta \cos\phi}{2(v_1 + w_x \cos\theta)} \right] \notag \\
	&= 0,
\end{align}
and
\begin{align}
	{\rm Tr}[\chi^{(4)}_z] &= \sum_j  \chi^{(4)}_{z;jj} \notag \\
	&= \sum_j \frac{e^2 \tau}{\hbar} \int \frac{d^3 k }{(2\pi)^3} v_z \partial_{k_j} m_j \frac{\partial f_0}{\partial \varepsilon} \notag \\
	&= -\frac{e^3 \tau}{\hbar^3} \frac{1 }{(2\pi)^3} \int_{0}^{\pi} d\theta \int_{0}^{2\pi} d\phi \left[  -\frac{v_1^2 \sin^2 \theta \sin\phi}{2(v_1 + w_x \cos\theta)} \right] \notag \\
	&= 0 .
\end{align}
Through similar derivation, we  find that
\begin{equation}
	{\rm Tr}[\chi^{(5)}_i] = {\rm Tr}[\chi^{(4)}_i].
\end{equation}

\subsection{Generic WP model}
Then, we derive analytical expressions of the PHE conductivity ${\rm Tr}[\chi^{(l)}_i]$ for a generic magnetic WP with arbitrary Chern number.
The Hamiltonian of  a generic magnetic WP with energy tilt can be described by the Eq. (7) in the main text, expressed as
\begin{eqnarray}\label{gwp}
  \mathcal H &=& w_xk_x + \mathcal H_0(\bm k) \nonumber \\
   &=& w_x k_x + \left[\begin{array}{cc}
		v_1 k_x & c_{+} k_{+}^n+c_{-} k_{-}^n \\
		c_{+} k_{-}^n+c_{-} k_{+}^n & -v_1 k_x
	\end{array}\right].
\end{eqnarray}
Here, without loss of generality, we assume the energy tilt is along $k_x$ direction,  $\mathcal H_0(\bm k)$ denotes a generic WP Hamiltonian without energy tilt.
As shown below, the PHE conductivities discussed here is solely determined by the Chern number of the $\mathcal H_0(\bm k)$, but is independent of the specific values of the  parameters in $\mathcal H_0(\bm k)$.
The dispersion for $\mathcal H$ is  $\varepsilon_{\bm k} = w_xk_x+\varepsilon_0$ with $\varepsilon_0$ the dispersion of   $\mathcal H_0(\bm k)$.

Again, we firstly study the expression of ${\rm Tr}[\chi^{(3)}_i]$, expressed as
\begin{align}\label{eq:chi3_1}
	{\rm Tr}[\chi^{(3)}_i] &= \sum_j  \chi^{(3)}_{i;jj}= -\frac{3e^3 \tau}{\hbar} \int \frac{d^3 k }{(2\pi)^3} v_i \left( \boldsymbol{\Omega}_{\bm{k}} \cdot \bm {v}_{\bm{k}} \right) \frac{\partial f_0}{\partial \varepsilon}  \notag \\
	&= \frac{3e^3\tau}{(2\pi)^3\hbar}\int v_i\Omega_\perp dS\int v_\perp\delta(\varepsilon - E_F) d k_\perp  \notag \\
	&= \frac{3e^3\tau}{(2\pi)^3\hbar^2}\oint_{\varepsilon = E_F} v_i\Omega_\perp dS,
\end{align}
where $\bm{v_k} \cdot \bm {\Omega_k} = v_\perp\Omega_\perp$ is adopted, $k_\perp$ is the  component of wave vector $\bm k$ normal to the Fermi surface,  and $v_\perp = \partial_{k_\perp}\varepsilon/\hbar$ is always positive for electron Fermi pocket.
The $x$-component velocity can be divided into two parts: $v_x = w_x/\hbar + \partial_{k_x}\varepsilon_0/\hbar = w_x/\hbar + v_{x,0}$, and the Eq. (\ref{eq:chi3_1}) can be further transformed into
\begin{align}\label{eq:chi3_2}
	{\rm Tr}[\chi_{x}^{(3)}] &= \frac{3e^3\tau}{(2\pi)^3\hbar^3}\oint_{\varepsilon = E_F} (w_x + \hbar v_{x,0})\Omega_\perp dS \notag \\
	&= 9\eta  w_x \mathcal C + \frac{3e^3\tau}{(2\pi)^3\hbar^2}\oint_{\varepsilon = E_F} v_{x,0}\Omega_\perp dS,
\end{align}
where the Chern number $\mathcal C$ is given by the integral $\oint_{\varepsilon = E_F} \Omega_\perp dS = 2\pi\mathcal C$.
Now, the task becomes to prove that the last term of Eq. (\ref{eq:chi3_2}) equals zero.

For a two-level system, it is convenient to map the system from $k$-space into $d$-space, where $\mathcal H_0(\bm k) = \bm d(\bm k) \cdot \bm s$ with ${\bm s}=(\sigma_z,\sigma_x, \sigma_y)$, and $d_1$, $d_2$, and $d_3$ denoting the coefficients of ${\bm s}$,
\begin{eqnarray}\label{dexpress}
  d_1 &=& v_1 k_x, \notag \\
  d_2 &=& (k_y +k_z)^{n/2} \left\{ c_{+} \cos[n\arg(k_y + ik_z)] + c_{-} \cos[n\arg(k_y - ik_z)] \right\}, \notag \\
  d_3 &=& -(k_y +k_z)^{n/2} \left\{ c_{+} \sin[n\arg(k_y + ik_z)] + c_{-} \sin[n\arg(k_y - ik_z)] \right\}
\end{eqnarray}
In the $d$-space, the Berry curvature takes a simple form of $\bm\Omega^d = -\frac{1}{2}\frac{\bm d}{d^3}$.
According to the expression of $\mathcal H $ [Eq. (\ref{gwp})], we also have  $v_{x,0} = \partial_{k_x}\sqrt{d_1^2 + d_2^2 +d_3^2}/\hbar =\frac{\partial d_1}{\partial k_x} d_1/(\hbar d)=v_1d_1/(\hbar d)$.
The mapping from $k$-space to $d$-space satisfies a useful property: $\Omega^k_\perp dS^k ={\bm \Omega^k \cdot d\bm S^k}={\rm Sign}({\cal{C}}){\bm \Omega^d \cdot d\bm S^d}={\rm Sign}({\cal{C}})\Omega^d_\perp dS^d$, which is proofed in Sec. \ref{detail}.
Then, the integral in the last term of  Eq. (\ref{eq:chi3_2}) becomes
\begin{align}\label{gauss}
	& \oint_{\varepsilon = E_F} v_{x,0}\Omega_\perp dS =
	-{\rm Sign}({\cal{C}})\times \frac{v_1}{\hbar} \oint_{d = d_F} \frac{d_1\bm d}{2d^4} \cdot d\bm S^d \notag \\
	&=  -{\rm Sign}({\cal{C}})\times \frac{v_1}{\hbar} \int \bm{\nabla_d} \cdot\left(\frac{d_1\bm d}{2d^4} \right) dV^d
	= 0,
\end{align}
as the gradient $\bm\nabla \cdot\left(\frac{d_i\bm d}{d^4} \right) = 0$ for $i =1, 2$, and $3$.
Thus, from Eq. (\ref{eq:chi3_2}), it is easy to know that
\begin{equation}
	{\rm Tr}[\chi^{(3)}_x]= 9\eta w_x\mathcal C.
\end{equation}
For ${\rm Tr}[\chi^{(3)}_{y(z)}]$, we have
\begin{align}\label{eq:chi3_y}
	{\rm Tr}[\chi_{y(z)}^{(3)}] &= \frac{3e^3\tau}{(2\pi)^3\hbar^2}\oint_{\varepsilon = E_F} v_{y(z)}(k,\theta, \phi)\Omega_\perp(k,\theta, \phi) dS,
\end{align}
According to  the WP Hamiltonian  $\mathcal H $ [Eq. (\ref{gwp})], it is easy to find out that  $v_{y(z)}(k,\theta, \phi)=-v_{y(z)}(k,\theta, \phi+\pi)$ and $\Omega_\perp(k,\theta, \phi)=\Omega_\perp(k,\theta, \phi+\pi)$. Thus, one has
\begin{equation}
	{\rm Tr}[\chi^{(3)}_y] = {\rm Tr}[\chi^{(3)}_z] = 0.
\end{equation}
Similarly, from  Eq. (3) in the main text, one has
\begin{eqnarray}
  {\rm Tr}[\chi^{(1)}_i] &=& -\frac{1}{3}{\rm Tr}[\chi^{(3)}_i], \\
  {\rm Tr}[\chi^{(2)}_i] &=& 3\eta w_x\mathcal C.
\end{eqnarray}

Next, we study the expressions of   ${\rm Tr}[\chi^{(4)}_i]$ and ${\rm Tr}[\chi^{(5)}_i]$, which are induced by  the orbital moment  $\bm m$.
Considering the definitions of orbital moment and Berry curvature
\begin{align}
	m_{n,\gamma} &= -e\hbar \epsilon_{\alpha\beta\gamma} \mathrm{Im} \sum_{m\neq n} \frac{v_{nm}^\alpha v_{mn}^\beta}{\varepsilon_n - \varepsilon_m}, \notag \\
	\Omega_{n,\gamma} &= -2\hbar^2 \epsilon_{\alpha\beta\gamma} \mathrm{Im} \sum_{m\neq n} \frac{v_{nm}^\alpha v_{mn}^\beta}{(\varepsilon_n - \varepsilon_m)^2}.
\end{align}
One finds that for the two-band WP model, these two quantities have a simple relation, i.e. $m_{n,\gamma} = \frac{e}{2\hbar}\Omega_{n,\gamma}(\varepsilon_n - \varepsilon_m)$. Therefore, we can rewrite  the expression of $\chi_{i; jj}^{(4)}$ as
\begin{align}
	\chi_{i; jj}^{(4)} =& -\frac{e^3\tau}{2\hbar} \int [d \bm{k} ] v_i (\varepsilon_c - \varepsilon_v)(\partial_{k_j} \Omega_{c,j})  \partial_{\varepsilon}f_k^0  -\frac{e^3\tau}{2} \int [d \bm{k} ] v_i \Omega_{c,j}(v_{c,j} - v_{v,j})  \partial_{\varepsilon}f_k^0,
\end{align}
where $\Omega_{c,j}$ is the Berry curvature of  conduction band, $\varepsilon_{c/v}$ denote the dispersion of conduction (valence) band, and  $v_{c(v),i} = \partial_{k_i}\varepsilon_{c(v)}/\hbar$.
We also define    $v_i\equiv v_{c,i}$ for simplification, as we focus on  the system  that is electron doped here.
Notice that in the region that does not have singularity, the gradient of Berry curvature is zero, \ie $\bm\nabla \cdot \bm\Omega_n = \sum_j\partial_{k_j} \Omega_{n,j}=0$.
For WP model (\ref{gwp}),  the velocity of conduction and valence band also has a simple relation $v_{c,j}({\bm k}) = -v_{v,j}({\bm k}) + 2\delta_{j,x}w_x/\hbar$. Then, we have
\begin{alignat}{2}\label{eq:chi4}
	{\rm Tr}[\chi^{(4)}_x] &=&& \sum_j \chi^{(4)}_{x;jj} \notag \\
	&=&& \frac{e^3\tau}{2\hbar} \oint_{\varepsilon = E_F} v_x (\varepsilon_c - \varepsilon_v)(\nabla \cdot \bm\Omega_c) dS
-e^3\tau \int [d \bm{k} ] v_x\sum_j\Omega_j (v_j - \delta_{j,x}w_x/\hbar) \partial_{\varepsilon}f_k^0 \notag \\
	&=&& -e^3\tau \int [d \bm{k} ] v_x\left( \bm {v}_{\bm{k}} \cdot \bm \Omega_{\bm{k}} \right) \partial_{\varepsilon}f_k^0  + \frac{e^3\tau w_x}{\hbar} \int [d \bm{k} ] v_x \Omega_x \partial_{\varepsilon}f_k^0  \notag \\
	&=&& -3\eta  w_x \mathcal C+ \frac{e^3\tau w_x}{\hbar} \int [d \bm{k} ] v_x \Omega_x \partial_{\varepsilon}f_k^0,
\end{alignat}
where the  result of  ${\rm Tr}[\chi^{(3)}_x]$ is used to obtain the last equation.
For the second term in the last line of Eq. [\ref{eq:chi4}], we have
\begin{align}\label{28}
	\int [d \bm{k} ] v_x \Omega_x \partial_{\varepsilon}f_k^0&=-\frac{1}{\left(2\pi\right)^{3} \hbar}\int dS dk_{\bot}  v_{x}\Omega_{x}\partial_{\varepsilon}f_{k}^{0} \notag\\
	&=-\frac{1}{\left(2\pi\right)^{3}\hbar}\int dS d\varepsilon \frac{1}{v_{\bot}} v_{x}\Omega_{x} \partial_{\varepsilon}f_{k}^{0}\notag\\
	&=\frac{1}{\left(2\pi\right)^{3}\hbar^2}\oint_{\varepsilon = E_F} dS\frac{v_{x}\Omega_{x}}{v_{\bot}}.
\end{align}
Since the velocity $\bm v$ at the Fermi surface  is always perpendicular to the Fermi surface, one knows that  $|\bm v|=v_{\bot}$, and the  $\bm v$  is parallel to the $d\bm S$. Thus,  $\frac{v_x}{v_{\bot}} = \frac{dS_{x}^k}{dS^k} $, as $dS=|d\bm S|$, leading to  $v_{x}\Omega_{x}/v_{\bot} dS	=	\Omega_{x}dS_{x}$.

\begin{figure}[t]
	\includegraphics[width=6.8 cm]{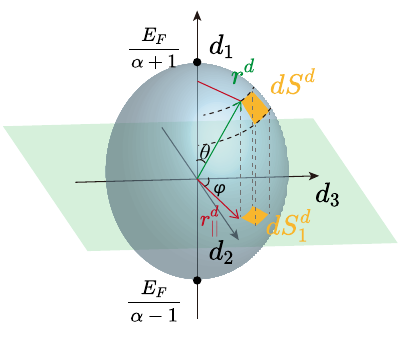}
	\caption{The Fermi surface of the tilted  WP in $d$-space.
}
	\label{figS1}
\end{figure}

It is also convenient to calculate Eq. (\ref{28}) in the $d$-space.
As demonstrated in  Sec. \ref{detail}, for the  WP Hamiltonian  $\mathcal H $ [Eq. (\ref{gwp})], we  have $\Omega_{x}^kdS_{x}^k=\Omega_{1}^ddS_{1}^d$, and then 
\begin{align}
	\int [d \bm{k} ] v_x \Omega_x \partial_{\varepsilon}f_k^0=\frac{1}{\left(2\pi\right)^{3}\hbar^2}\oint_{\varepsilon = E_F} dS\frac{v_{x}\Omega_{x}}{v_{\bot}}=\frac{1}{ \left(2\pi\right)^{3}\hbar^2}\oint_{\varepsilon = E_F} \Omega_{x}dS_{x}=\frac{1}{ \left(2\pi\right)^{3}\hbar^2}\oint_{\varepsilon = E_F} \Omega_{1}^ddS_{1}^d.
\end{align}
When $w_x\ne0$, the  WP Hamiltonian  $\mathcal H $ in $d$-space reads
\begin{align}
	\mathcal H=\alpha d_{1}+\bm d(\bm k) \cdot \bm s,
\end{align}
with $\alpha=w_x/v_1$. 
Then the Fermi surface of the WP in $d$-space is a ellipsoid (see Fig. \ref{figS1}), satisfying 
\begin{align}\label{eq:rd}
	\alpha d_{1}+\sqrt{d_{1}^{2}+(r_{||}^d)^2}=E_F,
\end{align}
with $r_{||}^d=\sqrt{d_{2}^{2}+d_{3}^{2}}$.
From Fig. \ref{figS1}, one notices that 
\begin{align}
	dS_{1}^{d}&=r_{||}^{d}d\phi^d dr^{d}=r_{||}^{d}\frac{\partial r_{||}^{d}}{\partial d_{1}}d\phi^d d\left(d_{1}\right),
\end{align}
where $r^d_{||}$ can be easily obtained from Eq.(\ref{eq:rd}) \ie $r_{||}^d = \sqrt{\left(E_F - \alpha d_1\right)^2 -d_1^2}$. The integration $\int \Omega_{1}^ddS_{1}^d$ then can be transformed as
\begin{align}
	\oint_{\varepsilon = E_F}\Omega_{1}^{d}dS_{1}^{d}=&\oint_{\varepsilon = E_F}\Omega_{1}^{d}r_{||}^{d}\frac{\partial r_{||}^{d}}{\partial d_{1}}d\phi^d d\left(d_{1}\right)\notag\\
	=&{\rm Sign}(\mathcal C)\int_0^{2n\pi }d\phi^d \int_{\frac{E_F}{\alpha-1}}^{\frac{E_F}{\alpha+1}}d\left(d_{1}\right)\Omega_{1}^{d}r_{||}^{d}\frac{\partial r_{||}^{d}}{\partial d_{1}}\notag\\
	=&\pi \mathcal C \int_{\frac{E_F}{\alpha-1}}^{\frac{E_F}{\alpha+1}}d\left(d_{1}\right)\frac{d_{1}}{\left(E_F-\alpha d_{1}\right)^{3}}\left[\left(\alpha^{2}-1\right)d_{1}-\alpha E_F\right]\notag\\
	=&2\pi \mathcal C \frac{\left(\alpha^{2}-1\right)\left(\alpha-\operatorname{arctanh}\alpha\right)}{\alpha^{3}}.
\end{align}
Finally, we have 
\begin{alignat}{2}
	{\rm Tr}[\chi^{(4)}_x] &=&& -3\eta  w_x \mathcal C+ \frac{e^3\tau w_x}{\hbar} \int [d \bm{k} ] v_x \Omega_x \partial_{\varepsilon}f_k^0\notag \\
&=&&-3\eta  w_x \mathcal C+\frac{e^3\tau w_x}{\hbar}\times \frac{1}{ \left(2\pi\right)^{3}\hbar^2}\times2\pi \mathcal C \frac{\left(\alpha^{2}-1\right)\left(\alpha-\operatorname{arctanh}\alpha\right)}{\alpha^{3}}\notag \\
&=&& (-2 - \beta) \eta  w_x \mathcal C
\end{alignat}
Besides, we have 
\begin{alignat}{2}
	{\rm Tr}[\chi^{(4)}_{y(z)}] &=&& \sum_j \chi^{(4)}_{y(z);jj} \notag \\
	&=&& -e^3\tau \int [d \bm{k} ] v_{y(z)}\left( \bm {v}_{\bm{k}} \cdot \bm \Omega_{\bm{k}} \right) \partial_{\varepsilon}f_k^0  + \frac{e^3\tau w_x}{\hbar} \int [d \bm{k} ] v_{y(z)} \Omega_x \partial_{\varepsilon}f_k^0  \notag \\
	&=&&  \frac{e^3\tau w_x}{\hbar} \int [d \bm{k} ] v_{y(z)} \Omega_x \partial_{\varepsilon}f_k^0\notag \\
	&=&& 0,
\end{alignat}
as $v_{y(z)}(k,\theta, \phi)=-v_{y(z)}(k,\theta, \phi+\pi)$ and $\Omega_x(k,\theta, \phi)=\Omega_x(k,\theta, \phi+\pi)$. 

For  $\chi^{(5)}_{i;jj}$, we can see that it exactly equal to $\chi^{(4)}_{i;jj}$, due to the following equations
\begin{eqnarray}
\chi^{(4)}_{i;jj} &=& 	-\frac{e^2 \tau}{(2\pi)^3\hbar} \int d^3 k  v_i (\partial_{k_j} m_j)  (\partial_{\varepsilon}f_k^0) \notag \\
	&=& -\frac{e^2 \tau}{(2\pi)^3\hbar^2}  \int dk_j dk_k \int dk_i \frac{\partial m_j}{\partial k_j} \frac{\partial f_k^0}{\partial k_i}  \notag \\
    &=& -\frac{e^2 \tau}{(2\pi)^3\hbar^2}  \int dk_j dk_k \int dk_i  [\frac{\partial(f_k^0\partial_{k_j} m_j)}{\partial k_i}-\frac{\partial^2 m_j}{\partial k_j \partial k_i} f_k^0] \notag \\
      &=& -\frac{e^2 \tau}{(2\pi)^3\hbar^2}  \int dk_j dk_k (f_k^0\partial_{k_j} m_j)|_{k_i=-\infty}^{k_i=\infty}+ \frac{e^2 \tau}{(2\pi)^3\hbar^2}  \int dk_j dk_k \int dk_i  \frac{\partial^2 m_j}{\partial k_j \partial k_i} f_k^0          \notag \\
      &=&\frac{e^2 \tau}{(2\pi)^3\hbar^2}  \int dk_j dk_k \int dk_i  \frac{\partial^2 m_j}{\partial k_j \partial k_i} f_k^0          \notag \\
       &=&-\frac{e^2 \tau}{(2\pi)^3\hbar^2}  \int dk_j dk_k \int dk_i  \frac{\partial m_j}{\partial k_i} \frac{\partial f_k^0}{\partial k_j}         \notag \\
	&=& -\frac{e^2 \tau}{(2\pi)^3\hbar} \int d^3k v_j (\partial_{k_i} m_j)  (\partial_{\varepsilon}f_k^0)\notag \\
	&=&\chi^{(5)}_{i;jj}.
\end{eqnarray}

\section{mapping from $k$-space to $d$-space} \label{detail}
In this section, we give the detail proof of the validity of the mapping from $k$-space to $d$-space in the integral.
The WP Hamiltonian in $k$-space is 
\begin{eqnarray}\label{k1}
  \mathcal H &=& w_x k_x + \left[\begin{array}{cc}
		v_1 k_x & c_{+} k_{+}^n+c_{-} k_{-}^n \\
		c_{+} k_{-}^n+c_{-} k_{+}^n & -v_1 k_x
	\end{array}\right],
\end{eqnarray}
and in  $d$-space is 
\begin{align}\label{d1}
	\mathcal H=\alpha d_{1}+\bm d(\bm k) \cdot \bm s.
\end{align}
The Chern number of this WP is ${\cal{C}}=n\times {\rm Sign}(v_1)$  for $|c_+|>|c_-|$ and is ${\cal{C}}=-n\times {\rm Sign}(v_1)$ for $|c_+|<|c_-|$.

\subsection{The first method}
Technically, the mapping is exactly  the integration by substitution, namely  
\begin{align}
	\int_{\varphi(k)}f({\bm d}) d{\bm d} =\int_k f(\varphi({\bm k}))\times |{\rm det}(J)| d{\bm k},
\end{align}
where substitution function $(d_1,...,d_n)=\varphi(k_1,...,k_n)$, and ${\rm det}(J)$  denotes the determinant of the Jacobian matrix $J$ of partial derivatives of $\varphi$ at point ${\bm k}$. 
Consider a 2D surface defined in  $k$-space $S^k$. The Berry curvature on $S^k$ is
\begin{align}
	\Omega_{\perp}^{k}=\left\langle \partial_{k_{1}}u|\partial_{k_{2}}u\right\rangle -\left\langle \partial_{k_{2}}u|\partial_{k_{1}}u\right\rangle,
\end{align}
where $k_1$ and $k_2$ are two  principal directions of surface $S^k$. Then we change it into $d$-space:
\begin{align}
	\left|\partial_{k_{1}}u\right\rangle =\frac{\partial d_{1}}{\partial k_{1}}\left|\partial_{d_{1}}u\right\rangle +\frac{\partial d_{2}}{\partial k_{1}}\left|\partial_{d_{2}}u\right\rangle, \notag \\
	\left|\partial_{k_{2}}u\right\rangle =\frac{\partial d_{1}}{\partial k_{2}}\left|\partial_{d_{1}}u\right\rangle +\frac{\partial d_{2}}{\partial k_{2}}\left|\partial_{d_{2}}u\right\rangle.
\end{align}
After some simple algebra, we arrive at
\begin{align}
	\Omega_{\perp}^{k}	&=\left(\frac{\partial d_{1}}{\partial k_{1}}\frac{\partial d_{2}}{\partial k_{2}}-\frac{\partial d_{1}}{\partial k_{2}}\frac{\partial d_{2}}{\partial k_{1}}\right)\left(\left\langle \partial_{d_{1}}u|\partial_{d_{2}}u\right\rangle -\left\langle \partial_{d_{2}}u|\partial_{d_{1}}u\right\rangle \right)\notag\\
	&=\left|\begin{array}{cc}
		\frac{\partial d_{1}}{\partial k_{1}} & \frac{\partial d_{1}}{\partial k_{2}} \\
		\frac{\partial d_{2}}{\partial k_{1}} & \frac{\partial d_{2}}{\partial k_{2}}
	\end{array}\right|\Omega_{\perp}^{d}\notag\\
	&={\rm det}(J) \Omega_{\perp}^{d},
\end{align}
with $J$ the Jacobian matrix.
Thus, one has  
\begin{eqnarray}
{\bm \Omega}^d\cdot  d{\bm S}^d &=& \Omega_{\perp}^d dS^d = \Omega_{\perp}^d\times |{\rm det}(J)| dS^k  = {\rm Sign}[{\rm det}(J)] \Omega_{\perp}^k  dS^k  = {\rm Sign}[{\rm det}(J)]{\bm \Omega}^k\cdot  d{\bm S}^k,
\end{eqnarray}
and 
\begin{eqnarray}
  \int_{\varphi(k)}{\bm \Omega}^d\cdot  d{\bm S}^d &=& \int_k{\rm Sign}[{\rm det}(J)] {\bm \Omega}^k\cdot  d{\bm S}^k.
\end{eqnarray}
For the WP Hamiltonian in Eqs. (\ref{k1}) and (\ref{d1}), we have 
\begin{eqnarray}
  \int_{\varphi(k)}{\bm \Omega}^d\cdot  d{\bm S}^d &=& \int_{\varphi(k)}\Omega_{\perp}^d dS^d=\int_{0}^{\pi} d\theta^d \int_{0}^{2n\pi} d\phi^d (r^d)^2 \Omega_{\perp}^d ={\rm Sign}({\cal{C}}) \int_k {\bm \Omega}^k\cdot  d{\bm S}^k.
\end{eqnarray}

Besides, for the  WP Hamiltonian in Eqs. (\ref{k1}) and (\ref{d1}), one notice that $d_1$ is only a function of $k_x$, while $d_2$ and $d_3$ are only functions of $k_y$ and $k_z$. This means that $dS_{x}^{k}=dk_{y}dk_{z}$ and $dS_{1}^{d}=dd_{2}dd_{3}$.
Thus, similar to the above derivation, we have 
\begin{eqnarray}
	\Omega_{x}^{k}&=&\left\langle \partial_{k_{y}}u|\partial_{k_{z}}u\right\rangle -\left\langle \partial_{k_{z}}u|\partial_{k_{y}}u\right\rangle, \nonumber \\
\left|\partial_{k_{y}}u\right\rangle &=&\frac{\partial d_{2}}{\partial k_{y}}\left|\partial_{d_{2}}u\right\rangle +\frac{\partial d_{3}}{\partial k_{y}}\left|\partial_{d_{3}}u\right\rangle, \ \ \ \ \ \ 
	\left|\partial_{k_{z}}u\right\rangle =\frac{\partial d_{2}}{\partial k_{z}}\left|\partial_{d_{2}}u\right\rangle +\frac{\partial d_{3}}{\partial k_{z}}\left|\partial_{d_{3}}u\right\rangle.
\end{eqnarray}
and 
\begin{align}
	\Omega_{x}^{k}	&=\left|\begin{array}{cc}
		\frac{\partial d_{2}}{\partial k_{y}} & \frac{\partial d_{2}}{\partial k_{z}} \\
		\frac{\partial d_{3}}{\partial k_{y}} & \frac{\partial d_{3}}{\partial k_{z}}
	\end{array}\right|\Omega_{1}^{d}={\rm det}(J') \Omega_{\perp}^{d},
\end{align}
with $J'$ the Jacobian matrix.
Thus, one has  
\begin{eqnarray}
\Omega_{1}^d dS^d_1 &=&   \Omega_{1}^d\times |{\rm det}(J')| dS_x^k  = {\rm Sign}[{\rm det}(J')] \Omega_{x}^k  dS_x^k.
\end{eqnarray}
According to $\mathcal H $ [Eq. (\ref{gwp})] and Eq. (\ref{dexpress}), ${\rm det}(J') = n^2(c_{-}^2 -c_{+}^2)(k_y^2 + k_z^2)^{n-1}$. Then, ${\rm Sign}[{\rm det}(J')]$ is $+1$ when $|c_{+}| < |c_{-}|$, and is $-1$, when $|c_{+}| > |c_{-}|$.

\subsection{The second method}
In two-band systems, we defined $d$-space by  $\mathcal H_0(\bm k) = \bm d(\bm k) \cdot \bm\sigma$ and $\bm d(\bm k)$ is a vector in $d$-space.
Consider a surface  in $k$-space, denoted as $S^k$ and its mapping in $d$-space is $S^d$. Then, consider four states on surface $S^k$: $\left|A\right\rangle$, $\left|B\right\rangle$,$\left|C\right\rangle$ and $\left|D\right\rangle$ which is very close to a point $\bm k_0$. The Berry phase $\gamma$ for the loop $ABCD$ is
\begin{align}
	\gamma=-\rm{Arg} \left\langle A|B\right\rangle \left\langle B|C\right\rangle \left\langle C|D\right\rangle \left\langle D|A\right\rangle.
\end{align}
Supporting the area surrounded by loop $ABCD$ is $\Delta S^k$ and the Berry curvature at $k_0$ in $S^k$ is $\Omega^k_\bot$, we can also express Berry phase as the integral of Berry curvature and in a enough small area, it  reduces to
\begin{align}
	\gamma \approx \Omega^k_\bot \cdot \Delta S^k.
\end{align}
Then we consider $\left|A\right\rangle$, $\left|B\right\rangle$,$\left|C\right\rangle$ and $\left|D\right\rangle$ in $d$-space. As the absolute value of Berry phase doesn't depend on parameter space, it is also $|\gamma|$. However, Berry curvature and surrounding area depend on parameter space, and respectively becomes $\Omega^d_\bot$ and $\Delta S^d$. Then, we  have
\begin{align}
	|\gamma| \approx \Omega^d_\bot \cdot \Delta S^d,
\end{align}
When the area is small enough, $\Delta S \rightarrow dS$ and we get
\begin{align}
	\left| {\bm\Omega}^d_\bot \cdot d {\bm S}^d \right| =\left| {\bm \Omega}^k_\bot \cdot d {\bm S}^k \right|.
\end{align}
This also means that when we proceed a mapping from on space to another, the flux of Berry curvature for a certain surface is unchanged.
